\documentclass[aip,apl,reprint]{revtex4-1}

\usepackage{graphicx}
\usepackage{amsmath}
\usepackage{mathcomp}	
\begin{document}

\title{Hybrid InAs nanowire-vanadium proximity SQUID}

\author{P.~Spathis$^{\ast}$,~S.~Biswas,~S.~Roddaro,~L.~Sorba,~F.~Giazotto$^{\ast}$,~and~F.~Beltram}
\affiliation{NEST, Istituto Nanoscienze-CNR and Scuola Normale Superiore, Piazza S. Silvestro 12, I-56126 Pisa, Italy}

\email{panayotis.spathis@sns.it,francesco.giazotto@sns.it}

\begin{abstract}
We report the fabrication and characterization of superconducting quantum interference devices (SQUIDs) based on InAs nanowires and vanadium superconducting electrodes. These mesoscopic devices are found to be extremely robust against thermal cycling and to operate up to temperatures of $\sim2.5$~K with reduced power dissipation. We show that our geometry allows to obtain nearly-symmetric devices with very large magnetic-field modulation of the critical current. All these properties make these devices attractive for on-chip quantum-circuit implementation.
\end{abstract}
\maketitle
A normal conductor (N) experiences superconducting correlations when in the vicinity of a superconductor (S). Owing to this $proximity$ $effect$ a gap $\varepsilon_G$ in the density of states (DOS) opens up in the N region, whose amplitude can be modulated by the phase of the superconducting order parameter~\cite{Zhou1998}. This can be readily done by applying a magnetic 
field with a loop geometry~\cite{PhysRevLett.73.2488, PhysRevLett.74.602,PhysRevLett.74.5268,PhysRevLett.100.197002,baselmans:2940}. Based on this principle, the superconducting quantum interference proximity transistor (SQUIPT)~\cite{Giazotto2010} was recently demonstrated. In this novel type of interferometer the phase-modulated DOS of a N metal in good contact with a superconducting loop is sensed by tunnel-coupled superconducting leads. This device operates similarly to a SQUID and shows good sensitivity~\cite{Giazotto2010}. DC proximity SQUIDs were also realized~\cite{PhysRevB.77.165408,vanDam2006,ref2006,graphene}. In all these examples, the SNS weak links are diffusive and belong to the long-junction limit. In this case, the N-region length is larger than the superconducting coherence length and governs the properties of the junctions. In this Letter, we present hybrid proximity DC SQUIDs in which weak link between vanadium (V) superconducting electrodes consists of an indium arsenide (InAs) nanowire (NW). We shall show that these weak links are in the intermediate-length junction regime and make it possible to reach rather high operating temperatures and sensitivity values.

\begin{figure}[t]
\includegraphics[width=\columnwidth,keepaspectratio]{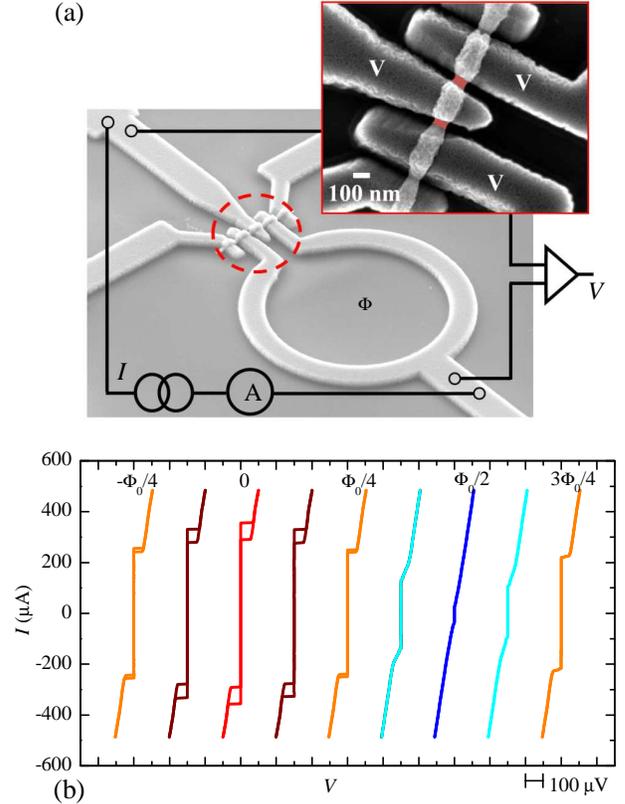}
\caption{\label{fig1} (color online) (a) SEM image of a typical device. The magnified area indicated by the dashed circle emphasizes the proximized region of the nanowire by the V contacts. b) Typical current-voltage ($I-V$) characteristics measured at 244~mK for perpendicular fields increasing in steps of $\Phi_0/8$, where $\Phi_0$ is the flux quantum. The curves are horizontally offset for clarity (dev.~C).}
\end{figure}

Highly-doped InAs NWs (with charge density $n=1.8\pm0.8\times10^{19}$~cm$^{-3}$) grown by chemical beam epitaxy were used for this study. NWs have diameters of $90\pm$10~nm and are around $2.5~\mu$m long. Figure~\ref{fig1}(a) shows a scanning electron micrograph (SEM) of a typical device obtained by e-beam lithography and subsequent e-beam evaporation of Ti/V (15/120 nm) in a UHV chamber. Transparency of the contacts is tuned by a passivation step of the InAs NW surface with a NH$_4$S$_x$ solution. The structure consists of a superconducting loop interrupted by two weak links obtained on a single NW, thus defining the SQUID. Weak links are colored in red in the magnified SEM image (inset of Figure~\ref{fig1}(a)). The additional outer V contacts on the NW are not used in this study. The length of the weak links is typically between $L=20$ and 50 nm while the diffusion constant of the NWs~\cite{2010arXiv1003.2140R} is $D=0.02$~m$^2$/s. Consequently the associated Thouless energy $E_{Th}=\hbar D/L^2$ is about one order of magnitude larger than the vanadium superconducting gap ($\Delta\sim 700~\mu$eV). Thus, the present links are not in the long-junction limit. The transport properties of these devices were measured in a filtered $^3$He refrigerator (two-stage RC and $\pi$-filters at 244~mK) using a standard 4-wire technique.

Below the transition temperature of the V contacts ($T_c\sim 4.6$~K), proximity effect can occur in the weak links. Figure~\ref{fig1}(b) shows several current-voltage ($I$-$V$) characteristics measured at 244~mK in the presence of a magnetic field applied perpendicularly to the plane of the ring. Data plotted were taken with one of the measured devices (dev.~C) and correspond to several magnetic field values. Dissipationless current flow is observed and demonstrates DC Josephson coupling in the device. This coupling is modulated by the applied field as expected from the interference between the split superconducting wavefunction along the arms of the loop. From these $I$-$V$s we can extract the critical current $I_c$, defined as the average of the maximum switching current for positive and negative bias, and the retrapping current $I_r$. Hysteresis is observed for critical currents above a threshold value (around 250~nA for dev.~C) and can be linked to electronic heating in the N region which occurs when the junction switches to the resistive branch~\cite{PhysRevLett.101.067002}. A similar behavior was also observed in InN NWs proximized by niobium electrodes~\cite{frielinghaus:132504}. The devices measured here exhibit critical currents ranging from $200$ to 350~nA. With N-state resistance values between $R_n=200$ and 300~$\Omega$, the product $eI_c R_n$ reaches roughly $100~\mu$eV$\sim0.14\Delta$. This energy is predicted to scale from $10.82 E_{Th}$ to $2.07 \Delta$ when going from the long ($E_{Th}\ll\Delta$) to the short ($\Delta\ll E_{Th}$) diffusive limit~\cite{ISI:A1970G761100037,PhysRevB.63.064502}. Therefore, our results indicate that the present junctions are in the {\it intermediate} regime. We believe this stems either from a residual Schottky barrier, or an oxide layer at the NW/V interfaces.
\begin{figure}[t]
\includegraphics[width=\columnwidth,keepaspectratio]{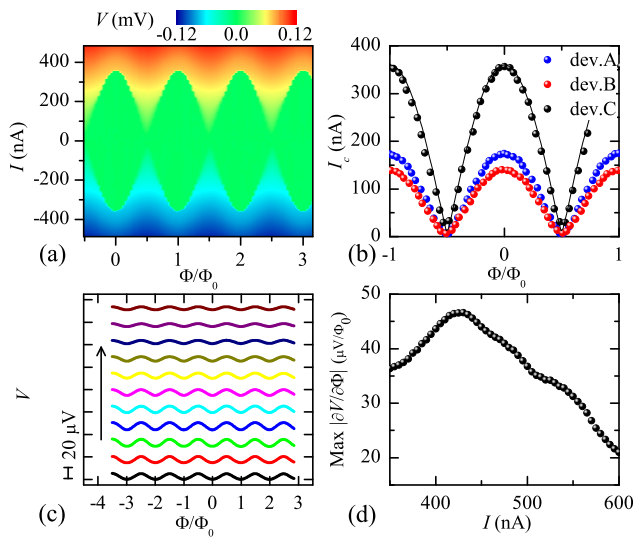}
\caption{\label{fig2} (color online) (a) Color plot of the voltage as a function of current bias and flux (dev.~C). (b) $I_c(\Phi)$ for three devices. A fit using Eq.~(\ref{fitSIS}) (full line) is plotted for dev.~C using $I_{c1}=175$ and $I_{c2}=185$~nA. (c) $V(\Phi)$ for biasing currents ranging from 355 to 655~nA in steps of 30~nA. The curves are horizontally offset for clarity (dev.~C). (d) Corresponding maximum values of $|\partial V/\partial \Phi|$ versus biasing current. All these measurements were performed at 244~mK.}
\end{figure}

Figure~\ref{fig2}(a) shows a color plot of the voltage measured as a function of magnetic field and biasing current when the latter is ramped from $0$ to both positive and negative values (dev.~C). In this way, we focus solely on the switching current. The Josephson coupling region displays zero-voltage drop and is shown in green color. A clear $\Phi_0$ periodicity can be observed, where $\Phi_0=h/2e$ is the flux quantum. In our devices, this period corresponds to a magnetic field of $\sim1.4$~Oe through an effective area of 14.6~$\mu$m$^2$, close to the geometrical area of 12.2~$\mu$m$^2$ inferred from the SEM image. Figure~\ref{fig2}(b) displays the corresponding critical current versus magnetic field together with the results of two other samples (dev.~A and B). The modulation depth of the critical current can be defined as $(I_c^{max}-I_c^{min})/I_c^{max}$ where $I_c^{max}$ and $I_c^{min}$ are the maximum and minimum critical current over one flux quantum. For all devices it is 95~\%. In addition, we find that the following relation holds: 
\begin{equation}\label{fitSIS}
I_c(\Phi)=\sqrt{I_{c1}^2+I_{c2}^2+2 I_{c1} I_{c2} \cos \left(\frac{2\pi \Phi}{\Phi_0}\right)},
\end{equation}
where $I_{c1}$ and $I_{c2}$ are the critical currents of each weak link [see fit in Fig.~\ref{fig2}(b)]. Equation~\ref{fitSIS} is expected to hold for junctions with conventional sinusoidal current-phase relations~\cite{clarke} (CPR) and negligible inductance. 
For fully-transparent short diffusive SNS junctions whose CPR is given by $I \propto \cos(\delta/2)\tanh^{-1}[\sin(\delta/2)]$ where $\delta$ is the phase difference across the junction~\cite{ISI:A1970G761100037,PhysRevB.66.184513}, the critical current of a SQUID can be modulated only up to 81~\% even for perfectly symmetric junctions~\cite{vijay:223112}. Our results thus show that a residual barrier lowering the transmissivity of the NW/V contacts can renormalize the Thouless energy and modify the CPR~\cite{PhysRevB.66.184513,PhysRevB.76.064514} towards the sinusoidal limit. In addition, a very small critical-current asymmetry of the two weak links is obtained from the fit ($I_{c1}/I_{c2}=0.95$). This stems from the fact that they are obtained by proximizing two close and homogeneous sections of a single NW and is a key advantage of our device. This almost-complete $I_c$ modulation also occurs because the external flux effectively modulates the phase difference on both weak links. To this end, we first reduced the self-inductance of the loop ($L_G$) by choosing the geometry so that $L_G\sim 2.5$~pH, and $2 L_G I_c/\Phi_0\sim 6.10^{-4}$. A second requirement is that the kinetic inductance of the superconducting loop ${\cal L}_S$ be negligible compared to that of the weak links 
${\cal L}_N$.~\cite{PhysRevLett.76.1402,*PhysRevB.66.220505} Their ratio is given by ${\cal L}_S/{\cal L}_N\sim(\varepsilon_G / \Delta) \times (R_n^S/R_n^N)$, where $R_n^S$ and $R_n^N$ are the normal-state resistance of the superconducting loop and the N part, respectively. The minigap amplitude is expected to reach the superconducting gap in the short diffusive limit, so that $\varepsilon_G/\Delta < 1$. Moreover, due to the presence of the InAs NW, we estimate $R_n^S\sim 0.3 R_n^N$. This ratio is indeed rather small and hinders phase-gradient establishment along the S loop~\cite{PhysRevLett.103.087003}, and allows therefore a proper phase biasing of the SQUID.

To operate the SQUID as a flux-to-voltage transformer, we current biased it above $I_c$. The critical-current periodicity thus transforms in a $\Phi_0$-periodic modulation of the voltage drop developed across the SQUID. Figure~\ref{fig2}(c) shows these modulations for biasing currents ranging between 355 and 655~nA. An important figure of merit of the SQUID is given by the maximum of the flux-to-voltage transfer function, max$|\partial V/\partial \Phi|$ plotted fig.~\ref{fig2}(d). A non-monotonic behavior is obtained similarly to the case of SQUIDs with conventional Josephson tunnel junctions~\cite{clarke}. The optimum biasing current for dev.~C is $\sim$425~nA, where the transfer function attains values as high as 45~$\mu$V/$\Phi_0$. Furthermore, the corresponding power dissipation is of the order of a few tens of pW making these devices attractive for applications such as quantum computing~\cite{vijay:223112}. By reducing the normal state resistance of the junctions one could in principle limit further the power dissipation.

\begin{figure}[t]
\includegraphics[width=\columnwidth,keepaspectratio]{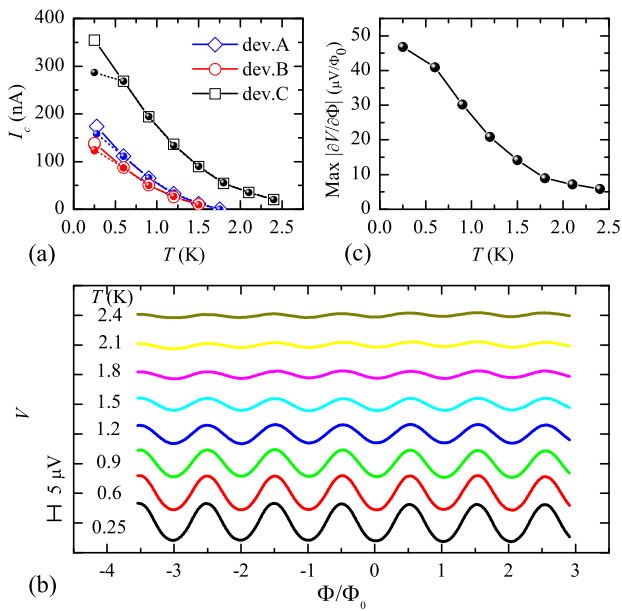}
\caption{\label{fig3} (color online) (a) Critical current dependence versus temperature for all devices (open symbols). The retrapping current is also plotted (solid dots). (b) $V(\Phi)$ at several bath temperatures for $I=425$~nA (curves are horizontally offset for clarity) and (c) corresponding maximum $|\partial V/\partial \Phi|$ for dev.~C.}
\end{figure}

Figure~\ref{fig3}(a) shows the switching (open symbols) and the retrapping current (solid dots) for the three devices as a function of temperature. The overall decrease of the critical current is similar for all devices and is the one typically observed for diffusive junctions in the intermediate-length limit~\cite{ISI:A1970G761100037}. In addition, the hysteresis disappears above 600 mK. This indicates one advantage of the present SNS junctions over conventional tunnel-junction SQUIDs since the former do not require a shunting resistance to avoid hysteresis.

DC Josephson coupling can be observed up to $\sim$2.5~K for dev.~C. The other devices studied showed DC Josephson effect up to lower temperatures since they present lower Thouless-energy values. Similar temperatures were also reported in previous studies on V/Cu/V junctions~\cite{garca:132508}. Figure~\ref{fig3}(b) shows the voltage modulation of dev.~C versus $\Phi$ for some representative bath temperatures. Oscillations of the transfer functions were observed up to 4.6~K (data not shown). Therefore, using NW-based weak links approaching the short diffusive limit allows us to benefit from the higher $T_c$ of vanadium. Figure~\ref{fig3}(c) shows the maximum of the transfer function as a function of temperature for dev.~C. At 1K, values exceeding $20~\mu$V$/\Phi_0$ are obtained.

In summary, we have presented the implementation and the characterization of DC SQUIDs based on InAs NWs and vanadium electrodes. The combination of these materials allowed us to obtain improved characteristics and higher operation temperatures. Using a single NW to define the two weak links of the interferometer enables us to obtain an almost-full modulation of the critical current. Importantly, these devices showed no degradation upon repeated thermal cycling. We believe these nanostructures may find application for quantum-computing applications where low noise and reduced dissipation are required.

We acknowledge F.~Taddei for fruitful discussions and partial financial support from NanoSciERA project ``Nanofridge'' and Fondazione Monte dei Paschi di Siena. 
\bibliographystyle{aipnum4-1}

\end{document}